\documentclass[symmetry,article,accept,pdftex,moreauthors]{Definitions/mdpi}
\usepackage{amsmath}
\usepackage{amsfonts}
\usepackage{amssymb,bm}
\usepackage{siunitx}
\usepackage{color}
\usepackage{braket}
\usepackage{graphics}

\firstpage{1}
\pubvolume{16}
\issuenum{3}
\articlenumber{320}
\pubyear{2024}
\copyrightyear{2024}
\datereceived{12 January 2024}
\daterevised{16 February 2024}
\dateaccepted{1 March 2024}
\datepublished{7 March 2024}
\hreflink{https://doi.org/ \linebreak 10.3390/sym16030320}

\Title{Nonequilibrium Casimir-Polder Force between Nanoparticles and
Graphene-Coated Silica Plate: Combined Effect of the Chemical
Potential and Mass Gap}

\TitleCitation{Nonequilibrium Casimir-Polder Force between nanoparticles and
Graphene-Coated Silica Plate: Combined Effect of the Chemical
Potential and Mass Gap}

\Author{{Galina L. Klimchitskaya} ${}^{1,2}$*\orcidA{},
Constantine C. Korikov${}^{3}$ and
Vladimir M. Mostepanenko ${}^{1,2,4,}$\orcidB{}
}

\AuthorNames{Galina L. Klimchitskaya, Constantine C. Korikov,
Vladimir M. Mostepanenko}

\AuthorCitation{Klimchitskaya, G.L.; Korikov, C.C.; Mostepanenko, V.M.}

\address{%
$^{1}$ \quad Central Astronomical Observatory at Pulkovo of the Russian Academy of Sciences, 196140 Saint Petersburg, Russia\\
$^{2}$ \quad Peter the Great Saint Petersburg
Polytechnic University, 195251 Saint Petersburg, Russia\\
$^{3}$ \quad Huawei Noah's Ark Lab, Krylatskaya str. 17, Moscow 121614, Russia \\
$^{4}$ \quad {Kazan Federal University}, 420008 Kazan, Russia}

\corres{Correspondence: g.klimchitskaya@gmail.com}

\abstract{The Casimir-Polder force between spherical nanoparticles and a
graphene-coated silica plate is investigated in situations out of
thermal equilibrium, i.e., with broken time-reversal symmetry.
The response of graphene coating to the electromagnetic field is
described on the basis of first principles of quantum electrodynamics
at nonzero temperature using the formalism of the polarization tensor
in the framework of the Dirac model. The nonequilibrium Casimir-Polder
force is calculated as a function of the mass-gap parameter, chemical
potential of graphene and temperature of the graphene-coated plate,
which can be both higher and lower than that of the environment. It is
shown that the force value increases with increasing chemical
potential, and this increase is more pronounced when the temperature
of a graphene-coated plate is lower than that of the environment.
The nonequilibrium force also increases with increasing temperature
of the graphene-coated plate. This increase is larger when the plate
is hotter than the environment. The effect is revealed that the
combined impact of the chemical potential $\mu$ and mass gap
$\Delta$ of graphene coating depends on the relationship between
$\Delta$ and 2$\mu$. If $2\mu>\Delta$ the magnitude of the
nonequilibrium force between nanoparticles and a cooled graphene-coated
plate becomes much larger than for a graphene coating with $\mu=0$.
The physical reasons explaining this effect are elucidated. Possible
applications of the obtained results are discussed.}


\begin{document}

\section{Introduction}

The two-dimensional (2D) hexagonal lattice of carbon atoms called
graphene \cite{1} has attracted particular attention of many experts in
both fundamental and applied physics because at low energies it
adequately fits the requirements of the Dirac model. What this means is
that graphene can be considered as consisting of light quasiparticles
described not by the Schr\"{o}dinger equation, as is customary in
condensed matter physics, but by the relativistic Dirac equation in
2D space and one-dimensional (1D) time, where the speed of light $c$
is replaced with the Fermi velocity $v_F\approx c/300$ \cite{2}.

For experts in fundamental physics, graphene is interesting because it
possesses unusual electrical, optical and mechanical properties which
offer outstanding possibilities to large-scale exploratory
investigations \cite{3,4}. At the same time, graphene suggests a wide
range of prospective applications in nanotechnology and material
science \cite{5,6}. Specifically, certain of the nanotechnological
devices rely upon interaction between nanoparticles and graphene-coated
plates for their operation \cite{7,8,9,10,11,12}.

The Casimir-Polder force \cite{13} acting between an electrically
neutral microscopic particle and a material plate is the relativistic
and quantum phenomenon. It is initiated by the zero-point and thermal
fluctuations of the electromagnetic field. In the state of thermal
equilibrium, when temperatures of the microparticle, plate, and of the
environment are equal, the Casimir-Polder force is described by the
Lifshitz theory \cite{14,15,16}. This theory expresses the force
value via the dynamic polarizability of a particle and the
frequency-dependent dielectric permittivity of the plate material.

There is an extensive literature devoted to calculations of the
Casimir-Polder force between atoms, molecules, nanoparticles and
diversified material surfaces, including graphene, by means of the
Lifshitz theory (see, e.g., Refs. \cite{16a,17,18,19,20,21,22,23,24,
25,26,27,28,29,30,31,32,33,34,35,35a,36,37,38,39,40,41,42}).
In doing so, the spatially nonlocal dielectric permittivities and
the 2D reflection coefficients of graphene sheets \cite{43} were
expressed via the polarization tensor of graphene in (2+1)
dimensional space-time \cite{44,45,46,47} in the framework of the
Dirac model.

In some instances, the condition of thermal equilibrium is violated
with the result that the Lifshitz theory becomes inapplicable. For
example, this may happen in the applications of graphene to solar
cells where the graphene-coated plate has different temperature
from that of nanoparticles and of the environment \cite{48,49}.
Although almost all physical theories, including the Lifshitz
theory, are symmetric relatively to the operation of time reversal,
in thermally nonequilibrium systems this symmetry is broken.

The generalization of the Lifshitz theory to the cases when the
condition of thermal equilibrium is violated was developed under
an assumption that each body in the physical system under
consideration is in the state of local thermal equilibrium
\cite{50,51,52,53}. As a result, it has become possible to calculate
the Casimir-Polder force in situations when the temperature of a
plate, a nanoparticle, an even of the environment are dissimilar.

The theory of nonequilibrium Casimir and Casimir-Polder forces was
further generalized to the case of nonplanar configurations
\cite{54,55,56,57,58} and for the plate materials with
temperature-dependent dielectric permittivities \cite{59,60,61,62,63}.
It was also confirmed experimentally by measuring the nonequilibrium
Casimir-Polder force between the Bose-Einstein condensate of ${}^{87}$Rb
atoms and a SiO$_2$ plate, which was heated as compared to the
environmental temperature\cite{64}.

Investigation of the nonequilibrium Casimir-Polder interaction between
spherical nanoparticles and graphene sheet using the formalism of the
polarization tensor has been started by Ref. \cite{65}. In that work,
the pristine graphene sheet freestanding in vacuum was considered.
This means that the crystal lattice of graphene was assumed to be
perfect with no foreign atoms and gapless quasiparticle spectrum.
According to the results of Ref. \cite{65}, the impact of
nonequilibrium conditions on the total Casimir-Polder force is the most
pronounced at short separations between a nanoparticle and a graphene
sheet. At these separations, the force may even change its sign and
become repulsive.

Real graphene sheets are, however, not pristine and, specifically, the
quasiparticles in graphene, although light, are not massless. This
leads to some nonzero mass gap $\Delta$ in the quasiparticle spectrum
\cite{3,66,67}. An impact of the nonzero mass gap on the nonequilibrium
Casimir-Polder interaction between nanoparticles and a freestanding in
vacuum graphene sheet was investigated in Ref. \cite{68} within the
formalism of the polarization tensor. It was shown that in this case
the nonequilibrium Casimir-Polder force remains attractive and it is
possible to control its value by varying the mass-gap parameter.

The freestanding graphene sheets are difficult, if not impossible, to
use in physical experiments and in nanodevices, where they are usually
deposited on some dielectric plates. An impact of the plate material
on the nonequilibrium Casimir-Polder force between nanoparticles and
a silica plate coated with gapped graphene was considered in
Ref. \cite{69}. It was shown that the presence of a silica plate
increases the magnitude of the nonequilibrium Casimir-Polder force. With
increasing mass-gap parameter, the magnitude of the nonequilibrium force
decreases and the impact of graphene coating on the force becomes smaller.

It should be noted that computations of the nonequilibrium Casimir-Polder
force acting on nanoparticles from real graphene sheet using the
polarization tensor are rather cumbersome. They are connected with
calculation of the multiple integrals of rapidly varying functions and
were realized on a supercomputer \cite{68,69}.

In addition to the nonzero mass gap, real graphene sheets are unavoidably
doped, i.e., contain some fraction of foreign atoms described by the
nonzero value of the chemical potential $\mu$ \cite{3,70,71}. This should
be taken into account in the theoretical computations intended for the
comparison with the measurement data as was successfully done in the
experiment on measuring the equilibrium Casimir force acting between an
Au-coated sphere and a graphene-coated SiO$_2$ plate \cite{72,73}.

The nonequilibrium Casimir interaction between two plates coated with
the gapless graphene sheets possessing the nonzero chemical potential
was investigated in Ref. \cite{73a} using the phenomenological,
spatially local, model for the response of graphene to the electromagnetic
field \cite{70,73b}.

In this paper, we investigate the Casimir-Polder force acting between
spherical nanoparticles and fused silica plate coated with gapped and
doped graphene sheet in situations out of thermal equilibrium using the
formalism of the polarization tensor. It is assumed that the temperature
of nanoparticles is the same as of the environment. However, the
temperature of the graphene-coated plate can be either lower or higher
than that of the environment. The spatially nonlocal response
of graphene to the fluctuating electromagnetic field is described
by the polarization tensor, which takes into account the
dependence of graphene properties on temperature, on the mass-gap
parameter, and on the chemical potential. In doing so, the explicit
expressions for the polarization tensor along the real frequency axis
are presented in the region of evanescent waves. The obtained results
for the nonequilibrium Casimir-Polder force are compared with those
for the equilibrium one in the same configuration.

It is shown that increasing of the chemical potential of graphene
coating leads to the monotonous increase of both the equilibrium and
nonequilibrium Casimir-Polder forces between a nanoparticle and a
graphene-coated plate. In doing so, for a heated graphene-coated
plate, the mass gap makes a lesser impact on the force if the chemical
potential takes a nonzero value. According to our results, for a
cooled graphene-coated plate the impact of chemical potential of
graphene coating on the nonequilibrium Casimir-Polder force
essentially depends on whether the inequality $2\mu>\Delta$ or
$2\mu<\Delta$ is satisfied. By and large, the impact of chemical
potential on the nonequilibrium force for a cooled graphene-coated
plate is stronger than for a heated one.

The paper is organized as follows. In Section 2, both the formalism
of the Lifshitz theory and the computational results for the equilibrium
Casimir-Polder force between nanoparticles and a silica plate coated
with gapped and doped graphene sheet are presented. Section 3 contains
the generalization of the Lifshitz theory of Casimir-Polder force for
the out-of-thermal-equilibrium conditions and the required expressions
for the polarization tensor of graphene along the real frequency axis
with due regard for different relationships between the mass gap and
chemical potential of graphene coating. In Section 4, the computational
results for the nonequilibrium Casimir-Polder force as a function of
separation are presented under different values of temperature, mass
gap and chemical potential of graphene coating. In Section 5, the reader
will find a discussion of the obtained results. Section 6 contains our
conclusions.

\section{Equilibrium Casimir-Polder Force between Nanoparticles and Silica Plate
Coated with Gapped and Doped Graphene}
\newcommand{\ve}{{\varepsilon}}
\newcommand{\ix}{{(i\xi_{E,l},k,\Delta,\mu,T_P)}}
\newcommand{\up}{{(u,k,\mu,T_p)}}
\newcommand{\qe}{{q_{E,l}(k)}}
\newcommand{\eq}{{q_{E,l}^{\,\varepsilon}(k)}}
\newcommand{\tq}{{\tilde{q}_{E,l}(k)}}
\newcommand{\feq}{{F_{\rm eq}^{\rm SiO_2}}}
\newcommand{\nfeq}{{F_{\rm neq}^{\rm SiO_2}}}


We consider the spherical nanoparticles of radius $R$ spaced above large,
graphene-coated fused silica (SiO${}_2$) plate at height $a\gg R$.
The graphene coating is characterized by the mass gap $\Delta$ and
chemical potential $\mu$. The fused silica plate chosen for computations
is the most typical substrate used as a supporter for graphene
\cite{72,73,74,75,76,77}. All below equations are, however, applicable to
any material substrate.

The temperature of the environment is $T_E=300~$K. In this section, the
temperature of graphene-coated plate, $T_p$, is equal to $T_E$:
$T_p=T_E=300~$K. Nanoparticle temperature is assumed to be equal to
$T_E$ throughout the paper. We also assume that at all temperatures $T$
considered below the nanoparticle radius satisfies the condition
$R\ll \hbar c/(k_BT)$, where $\hbar$ is the reduced Planck constant,
$c$ is the speed of light, and $k_B$ is the Boltzmann constant.
At $T=T_E$ this condition reduces to the inequality $R\ll7.6~\upmu$m.
When it is satisfied, the dynamic polarizability of nanoparticle
$\alpha(\omega)$ at all frequencies $\omega$ contributing to the
Casimir-Polder force is approximately equal to the static value
$\alpha(0)$ \cite{57}.

In the situation of thermal equilibrium $T_p=T_E$, the Casimir-Polder
force between a nanoparticle and a graphene-coated plate is given
by the Lifshitz formula written in terms of the discrete Matsubara
frequencies \cite{14,15,16,78,79}. Keeping in mind application
of the same notations in the next section devoted to the situation
out of thermal equilibrium, in this section we continue to denote
the plate and environment temperatures as $T_p$ and $T_E$ but imply
that  $T_p=T_E=300~$K. Then, the equilibrium Casimir-Polder force
between a nanoparticle and a graphene-coated SiO${}_2$ plate is given
by \cite{14,15,16,78,79}
\begin{linenomath}
\begin{eqnarray}
&&
\feq(a,\Delta,\mu,T_E,T_p)=-\frac{2k_BT_E\alpha(0)}{c^2}\sum_{l=0}^{\infty}
{\vphantom{\sum}}^{\prime}\int\limits_{0}^{\infty}k\,dk\,e^{-2a\qe}
\label{eq1}\\
&&~~\times
\left\{\left[2q_{E,l}^2(k)c^2-\xi_{E,l}^2\right]R_{\rm TM}\ix-
\xi_{E,l}^2R_{\rm TE}\ix\right\}.
\nonumber
\end{eqnarray}
\end{linenomath}
Here, $k$ is the magnitude of the wave vector projection on the plane
of graphene-coated plate, the Matsubara frequencies are
$\xi_{E,l}=2\pi k_BT_El/\hbar$ with $l=0,\,1,\,2,\,\ldots\,$, the prime
on the summation sign means that the term with $l=0$ is divided by 2,
and

\begin{equation}
\qe=\left(k^2+\frac{\xi_{E,l}^2}{c^2}\right)^{1/2}.
\label{1a}
\end{equation}

The reflection coefficients on the graphene-coated plate for the
electromagnetic fluctuations with transverse magnetic (TM) and
transverse electric (TE) polarizations are expressed via the dielectric
permittivity of the plate material $\ve(\omega)$ and the polarization
tensor of graphene $\Pi_{\beta\gamma}(\omega,k,\Delta,\mu,T_p)$
calculated at the pure imaginary Matsubara frequencies \cite{38,72,73,80}

\begin{adjustwidth}{-\extralength}{0cm}
\begin{eqnarray}
&&
R_{\rm TM}\ix=\frac{\hbar k^2[\ve(i\xi_{E,l})\qe-\eq]+
\qe\eq\Pi_{00}\ix}{\hbar k^2[\ve(i\xi_{E,l})\qe+\eq]+
\qe\eq\Pi_{00}\ix},
\nonumber \\
&&
R_{\rm TE}\ix=\frac{\hbar k^2[\qe-\eq]-\Pi\ix}{\hbar k^2
[\qe+\eq]+\Pi\ix}.
\label{eq2}
\end{eqnarray}
\end{adjustwidth}

Here,
\begin{linenomath}
\begin{equation}
\eq=\left[k^2+\ve(i\xi_{E,l})\frac{\xi_{E,l}^2}{c^2}\right]^{1/2}
\label{eq2a}
\end{equation}
\end{linenomath}
and the following combination of the components of the polarization tensor is
introduced:
\begin{linenomath}
\begin{equation}
\Pi\ix\equiv k^2\Pi_{\beta}^{\,\beta}\ix-q_{E,l}^2(k)\Pi_{00}\ix,
\label{eq3}
\end{equation}
\end{linenomath}
where a summation over the repeated index $\beta=0,\,1,\,2$
is implied.

For numerical computations of the equilibrium Casimir-Polder force (\ref{eq1})
between a nanoparticle and a graphene-coated substrate, one needs the values
of the static polarizability of a nanoparticle, $\alpha(0)$, of the dielectric
permittivity of plate material, $\ve(i\xi_{E,l})$, and of the components of the
polarization tensor of graphene, $\Pi_{00}\ix$ and $\Pi\ix$.

The static polarizability of spherical nanoparticles made of dielectric material
is \cite{57} $\alpha(0)=R^3[\ve(0)-1]/[\ve(0)+2]$. Thus, for nanoparticles made
of high-resistivity Si one obtains $\ve(0)=3.81$ and $\alpha(0)=0.484R^3$.
For metallic nanoparticles, $\ve(0)=\infty$ and $\alpha(0)=R^3$.

The dielectric permittivity of SiO${}_2$ at the pure imaginary Matsubara
frequencies is obtained from the optical data for the imaginary part of
$\ve(\omega)$ \cite{81} with the help of the Kramers-Kronig relation.
The obtained values were repeatedly used in calculations of the Casimir and
Casimir-Polder forces \cite{78}. There are also analytic expressions for the
dielectric permittivity of SiO${}_2$ along the imaginary frequency axis
\cite{81a,81b}.

Below we present the components of the polarization tensor of gapped and
doped graphene at the Matsubara frequencies found in the framework of the
Dirac model \cite{44,45,46,47}. As mentioned in Section~1, this model is
applicable at low energies. In Ref.~\cite{82} the upper boundary of its
application region is estimated as 3~eV. Thus, one should consider not
too small separations $a$ between the nanoparticle and the graphene-coated
plate in order that all energies giving the major contribution to the
Casimir-Polder interaction lie below this boundary.

It has been known  that the characteristic energy of the Casimir-Polder
interaction is equal to $\hbar c/(2a)$ \cite{78,83}. This energy is less
than 1~eV at all separation distances $a>100~$nm. That is why, at separations
exceeding,
e.g., 200~nm one can reliably use the polarization tensor of graphene
derived in the framework of the Dirac model. This conclusion was confirmed by
measurements of the Casimir force between an Au-coated sphere and a
graphene-coated SiO${}_2$ plate, which were found in a very good agreement
with theoretical predictions using the polarization tensor of graphene
\cite{72,73}.

The explicit expression for the component $\Pi_{00}$ of the polarization
tensor is given by \cite{38,72,73}
\begin{linenomath}
\begin{eqnarray}
&&
\Pi_{00}\ix=\frac{\alpha\hbar k^2}{\tq}\Psi(D_{E,l})+
\frac{4\alpha\hbar c^2\tq}{v_F^2}\int\limits_{D_{E,l}}^{\infty}du\,w_{E,l}\up
\nonumber\\
&&~~~\times\left[1-{\rm Re}\frac{1- u^2+2i\gamma_{E,l}u}{\left(1-
u^2+2i\gamma_{E,l}u+D_{E,l}^2-\gamma_{E,l}^2D_{E,l}^2\right)^{1/2}}\right],
\label{eq4}
\end{eqnarray}
\end{linenomath}
where $\alpha=e^2/(\hbar c)$ is the fine structure constant and the following
notations are introduced

\begin{eqnarray}
&&
\Psi(x)=2\left[x+(1-x^2)\arctan\frac{1}{x}\right], \qquad
D_{E,l}=\frac{\Delta}{\hbar c\tq},
\nonumber \\
&&
\tq=\frac{1}{c}\sqrt{v_F^2k^2+\xi_{E,l}^2}, \qquad
\gamma_{E,l}=\frac{\xi_{E,l}}{c\tq},
\label{eq5} \\
&&
w_{E,l}\up=\sum_{\kappa=\pm 1}\left(e^{B_{E,l}u+\kappa\frac{\mu}{k_BT_p}}
+1\right)^{-1}, \qquad
B_{E,l}=\frac{\hbar c \tq}{2k_BT_p}.
\nonumber
\end{eqnarray}

In a similar way, the explicit expression for the combination of the
components of the polarization tensor $\Pi$  takes the form \cite{38,72,73}

\begin{eqnarray}
&&
\Pi\ix={\alpha\hbar k^2}{\tq}\Psi(D_{E,l})-
\frac{4\alpha\hbar \tq\xi_{E,l}^2}{v_F^2}\int\limits_{D_{E,l}}^{\infty}du\,w_{E,l}\up
\nonumber\\
&&~~~\times\left[1-{\rm Re}\frac{(1+i\gamma_{E,l}^{-1}u)^2+
(\gamma_{E,l}^{-2}-1)D_{E,l}^2}{\left(1-
u^2+2i\gamma_{E,l}u+D_{E,l}^2-\gamma_{E,l}^2D_{E,l}^2\right)^{1/2}}\right].
\label{eq6}
\end{eqnarray}

Computations of the equilibrium Casimir-Polder force between nanoparticles and
graphene-coated SiO${}_2$ plate were performed by Eqs.~(\ref{eq1}), ( \ref{eq2}),
(\ref{eq4}), and (\ref{eq6}). Since the force values quickly decrease with
increasing separation, they were normalized either to the Casimir-Polder force
$F_0$ between a nanoparticle and an ideal metal plane at zero temperature or
to the classical limit $F_{\rm cl}$ reached in the same configuration, considered
at temperature $T_E$ at large separations \cite{78,83}

\begin{equation}
F_0(a)=-\frac{3\hbar c}{2\pi a^5}\,\alpha(0), \qquad
F_{\rm cl}(a,T_E)=-\frac{3k_BT_E}{4a^4}\,\alpha(0).
\label{eq7}
\end{equation}

In Figure~\ref{fg1}, the computational results for (a) $F_{\rm eq}^{\rm SiO_2}/F_0$
and (b) $F_{\rm eq}^{\rm SiO_2}/F_{\rm cl}$ are presented as the function of separation
for a graphene coating with the mass gap $\Delta=0.2~$eV and chemical potential
$\mu=0$, 0.075, and 0.15~eV (lines 1, 2, and 3, respectively). As is seen in
Figure~\ref{fg1}, the force magnitude increases with increasing chemical
potential as it should be. At all separations considered the magnitude of
$F_{\rm eq}^{\rm SiO_2}$ decreases with separation slower than $|F_0|$.
At $a>1~\upmu$m, $|F_{\rm eq}^{\rm SiO_2}|$ decreases with separation
faster than $|F_{\rm cl}|$.
\begin{adjustwidth}{-\extralength}{0cm}
\begin{figure}[H]
\centerline{\hspace*{-2.7cm}
\includegraphics[width=7.in]{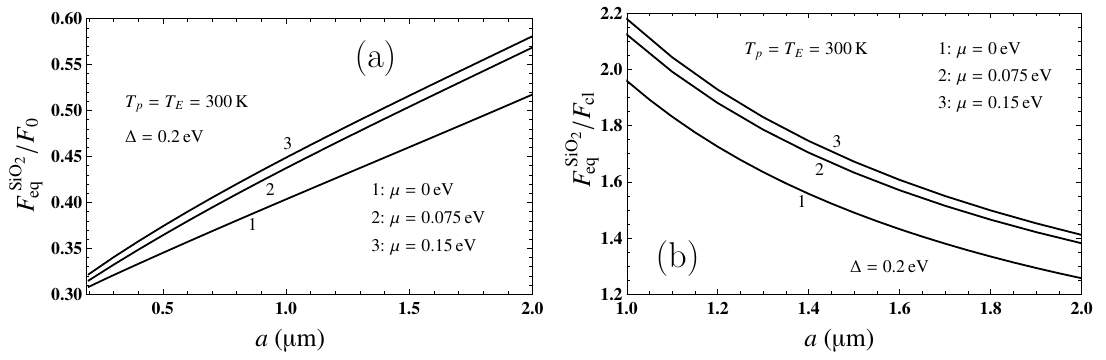}}
\caption{\label{fg1}
The equilibrium Casimir-Polder force between a nanoparticle and a graphene-coated
SiO${}_2$ plate at $T_p=T_E=300~$K normalized to (a) the zero-temperature Casimir-Polder
force from an ideal metal plane and (b) classical limit of the same force at
$T_E=300~$K is shown as the function of separation
for a graphene coating with the mass gap $\Delta=0.2~$eV and chemical potential
$\mu=0$, 0.075, and 0.15~eV  by the lines labeled 1, 2, and 3, respectively.
}
\end{figure}
\end{adjustwidth}

\begin{figure}[H]
\centerline{\hspace*{-2.7cm}
\includegraphics[width=5.in]{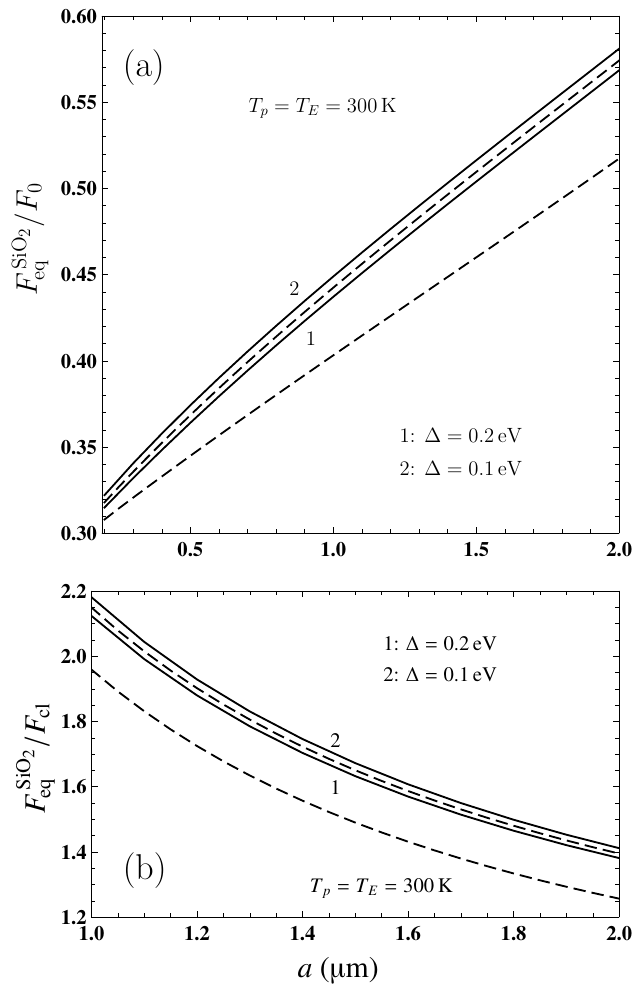}}
\caption{\label{fg2}
The equilibrium Casimir-Polder force between a nanoparticle and a graphene-coated
SiO${}_2$ plate at $T_p=T_E=300~$K normalized to (a) the zero-temperature Casimir-Polder
force from an ideal metal plane and (b) classical limit of the same force at
$T_E=300~$K is shown as the function of separation by the two pairs solid and dashed
lines. In the pair labeled 1, the mass-gap parameter $\Delta=0.2~$eV for both lines,
whereas the chemical potential $\mu=0.075~$eV for the solid line and 0 for the dashed
line. In the pair labeled 2, $\Delta=0.1~$eV for both lines with the same corresponding
values of $\mu$ as in the pair 1.
}
\end{figure}

Next, let us consider the impact of the mass-gap parameter on the Casimir-Polder
force. In Figure~\ref{fg2}, the computational results for (a) $F_{\rm eq}^{\rm SiO_2}/F_0$
and (b) $F_{\rm eq}^{\rm SiO_2}/F_{\rm cl}$ are presented as the function of separation
by the pairs solid and dashed lines labeled 1 and 2.
In pair 1, the mass-gap parameter $\Delta=0.2~$eV for both lines,
whereas the chemical potential $\mu=0.075~$eV for the solid  and 0 for the dashed
line. In pair 2, the mass-gap parameter $\Delta=0.1~$eV for both lines
with the same values of $\mu$, as in pair 1, for the solid and dashed lines.

As is seen in Figure~\ref{fg2}, the force magnitude increases with decreasing value
of the energy gap. This means that the chemical potential and the energy gap of
graphene coating act on the Casimir-Polder force in the opposite directions.
It is seen also that if $\mu\neq 0$ (see the solid lines) the dependence of the
force value on $\Delta$ is much weaker than for graphene coating with $\mu=0$
(see the dashed lines).

Note that numerical results presented in Figures~\ref{fg1} and \ref{fg2}
do not depend on $\alpha(0)$, i.e., on the nanoparticle radius. The absolute
values of  $F_{\rm eq}^{\rm SiO_2}$ can be obtained from
Figures~\ref{fg1} and \ref{fg2} by using Eq.~(\ref{eq7}) and the values of $\alpha(0)$
indicated above.

\section{Nonequilibrium Casimir-Polder Force between Nanoparticle and Plate
Coated with Gapped and Doped Graphene: General Formalism}
\newcommand{\po}{{(\omega,k,\Delta,\mu,T_p)}}
\newcommand{\zo}{{(\omega,k,\Delta,\mu,0)}}
\newcommand{\Ww}{{w(u,\omega,k,\mu,T_p)}}
\newcommand{\Tw}{{\widetilde{w}(u,\omega,k,\mu,T_p)}}
\newcommand{\aEP}{{(a,\Delta,\mu,T_E,T_p)}}
\newcommand{\oq}{{q(\omega,k)}}
\newcommand{\wq}{{q^{\,\varepsilon}(\omega,k)}}
\newcommand{\kp}{{p(\omega,k)}}
\newcommand{\gt}{{\widetilde{\gamma}(\omega,k)}}
\newcommand{\ktq}{{\tilde{q}(\omega,k)}}

Now we consider the situation out of thermal equilibrium when the temperature of
nanoparticles is equal to $T_E$, i.e., is the same as of the environment, whereas
the temperature of the graphene-coated plate $T_p$ can be either lower or higher
than $T_E$.

In the generalization of the Lifshitz theory to out-of-thermal-equilibrium conditions,
the nonequilibrium force is usually presented as the sum of two terms, one of which
is akin the equilibrium force by its form and another one is a truly nonequilibrium
term \cite{53,55}. In the literature, different forms of these terms are contained
leading to the same total result. Below, we represent the nonequilibrium
Casimir-Polder force following Refs.~\cite{65,68,69}

\begin{equation}
\nfeq\aEP=\feq\aEP+F_r^{\rm SiO_2}\aEP.
\label{eq8}
\end{equation}

Here, $\feq$ was already defined in Eq.~(\ref{eq1}) as an equilibrium Casimir-Polder
force, but now it has another meaning. In Eq.~(\ref{eq1}), it was assumed that
$T_p=T_E=300~$K, whereas now  $T_p\neq T_E$ and, thus, $\feq$, although resembles
the equilibrium force by its form, describes some part of the effects of
nonequilibrium.

The truly nonequilibrium contribution to Eq.~(\ref{eq8}) is given by \cite{65,68,69}
\begin{linenomath}
\begin{eqnarray}
&&
F_r^{\rm SiO_2}(a,\Delta,\mu,T_E,T_p)=\frac{2\hbar\alpha(0)}{\pi c^2}
\int\limits_{0}^{\infty}d\omega\,\Theta(\omega,T_E,T_p)
\int\limits_{\omega/c}^{\infty}k\,dk\,e^{-2a\oq}
\label{eq9}\\
&&~~\times{\rm Im}
\left\{\left[2q^2(\omega,k)c^2+\omega^2\right]R_{\rm TM}\po+
\omega^2R_{\rm TE}\po\right\},
\nonumber
\end{eqnarray}
\end{linenomath}
where

\begin{equation}
\Theta(\omega,T_E,T_p)=\left(e^{\frac{\hbar\omega}{k_BT_E}}-1\right)^{-1}-
\left(e^{\frac{\hbar\omega}{k_BT_p}}-1\right)^{-1}, \qquad
\oq=\left(k^2-\frac{\omega^2}{c^2}\right)^{1/2}.
\label{eq10}
\end{equation}

The reflection coefficients in Eq.~(\ref{eq9}) are similar to those in Eq.~(\ref{eq2}),
but now they are defined along the real frequency axis
\begin{linenomath}
\begin{adjustwidth}{-\extralength}{0cm}
\begin{eqnarray}
&&
R_{\rm TM}\po=\frac{\hbar k^2[\ve(\omega)\oq-\wq]+
\oq\wq\Pi_{00}\po}{\hbar k^2[\ve(\omega)\oq+\wq]+
\oq\wq\Pi_{00}\po},
\nonumber \\
&&
R_{\rm TE}\po=\frac{\hbar k^2[\oq-\wq]-\Pi\po}{\hbar k^2
[\oq+\wq]+\Pi\po},
\label{eq11}
\end{eqnarray}
\end{adjustwidth}
\end{linenomath}
where
\begin{eqnarray}
&&
q^{\,\ve}(\omega,k)=\left[k^2-\ve(\omega)\frac{\omega^2}{c^2}\right]^{1/2},
\nonumber\\
&&
\Pi\po=k^2\Pi_{\beta}^{\,\beta}\po-q^2(\omega,k)\Pi_{00}\po.
\label{eq12}
\end{eqnarray}

It is seen that Eq.~(\ref{eq11}) is obtained from Eq.~(\ref{eq2}) simply by putting
$i\xi_{E,l}=\omega$. However, the components of the polarization tensor (\ref{eq4})
and (\ref{eq6}) are rather complicated functions. They should be analytically
continued from the imaginary to real frequency axis of the plane of complex
frequency which is a nontrivial procedure.

The advantage of representation (\ref{eq8}) is that the truly nonequilibrium
contribution to it (\ref{eq9}) contains an integration on only the evanescent waves.
For these waves $k>\omega/c$ and the quantity $\oq$ in the power  of the exponent
under the integral is real.

The analytic continuation of the polarization tensor to the real frequency axis under
the condition $k>\omega/c$ is, nevertheless, rather involved. It takes different
forms in the so-called plasmonic region \cite{84}\begin{linenomath}
\begin{equation}
\frac{\omega}{c}<k\leqslant\frac{\omega}{v_F}\approx 300\frac{\omega}{c}
\label{eq13}
\end{equation}
\end{linenomath}
and in the region

\begin{equation}
300\frac{\omega}{c}\approx\frac{\omega}{v_F}<k<\infty.
\label{eq14}
\end{equation}

We begin with the plasmonic region (\ref{eq13}). The analytic continuation of the
polarization tensor to this region  of the real frequency axis for a pristine and
gapped graphene was performed in Refs.~\cite{65,68,69}. Here, we present the obtained
results in a more convenient analytic form and generalize them for the case of gapped
graphene with nonzero chemical potential.

As was shown in Ref.~\cite{46}, the form of the sought for analytic continuation
depends on whether $\hbar cp<\Delta$ or $\hbar cp\geqslant\Delta$
where $p\equiv\kp=\sqrt{\omega^2-v_F^2k^2}/c$. If the inequality $\hbar cp(\omega,k)<\Delta$ is satisfied, the results of Ref.~\cite{46} for $\Pi_{00}$ generalized for the case
$\mu\neq 0$ can be identically represented in the following form:
\begin{linenomath}
\begin{eqnarray}
&&
\Pi_{00}\po=-\frac{2\alpha k^2}{cp^2(\omega,k)}\,\Phi_1(\omega,k,\Delta)+
\frac{4\alpha\hbar c^2\kp}{v_F^2}\int\limits_{\widetilde{D}}^{\infty}
du\,\Tw
\nonumber\\
&&~~~\times
\left[1-\frac{1}{2c\kp}\sum_{\lambda=\pm 1}\lambda B_1(c\kp u+\lambda\omega)\right],
\label{eq15}
\end{eqnarray}
\end{linenomath}
where
\begin{linenomath}
\begin{eqnarray}
&&
\Phi_1(\omega,k,\Delta)=\Delta-\hbar c\kp\left[1+\frac{\Delta^2}{\hbar^2c^2p^2(\omega,k)}
\right]\,{\rm arctanh}\frac{\hbar c\kp}{\Delta},
\nonumber\\
&&
\Tw=\sum_{\kappa=\pm 1}\left[e^{\widetilde{B}(\omega,k,T_p)u+\kappa\frac{\mu}{k_BT_p}}
+1\right]^{-1}\!\!\!,
\nonumber\\
&&
\widetilde{D}\equiv\widetilde{D}(\omega,k,\Delta)=\frac{\Delta}{\hbar c\kp}, \qquad
\widetilde{B}(\omega,k,T_p)=\frac{\hbar c\kp}{2k_BT_p},
\nonumber \\
&&
B_1(x)=\frac{x^2-v_F^2k^2}{[x^2-v_F^2k^2A(\omega,k,\Delta)]^{1/2}}, \qquad
A(\omega,k,\Delta)=1-\frac{\Delta^2}{\hbar^2c^2p^2(\omega,k)} .
\label{eq16}
\end{eqnarray}
\end{linenomath}
In the region $\hbar c\kp<\Delta$ under consideration $A(\omega,k,\Delta)<0$ and
the quantity $\Pi_{00}$ is real.

Using the same notations, for $\Pi$ one obtains from Ref.~\cite{46}
\begin{linenomath}
\begin{eqnarray}
&&
\Pi\po=\frac{2\alpha k^2}{c}\,\Phi_1(\omega,k,\Delta)+
\frac{4\alpha\hbar\omega^2\kp}{v_F^2}\int\limits_{\widetilde{D}}^{\infty}
du\,\Tw
\nonumber\\
&&~~~\times
\left[1-\frac{c\kp}{2\omega^2}\sum_{\lambda=\pm 1}\lambda B_2(c\kp u+\lambda\omega)\right],
\label{eq17}
\end{eqnarray}
\end{linenomath}
where
\begin{linenomath}
\begin{equation}
B_2(x)=\frac{x^2+v_F^2k^2[1-A(\omega,k,\Delta)]}{[x^2-v_F^2k^2A(\omega,k,\Delta)]^{1/2}}.
\label{eq18}
\end{equation}
\end{linenomath}
The quantity $\Pi$ is also the real function.

Now we consider the region of $\omega$ and $k$ plane where the inequality
$\hbar cp(\omega,k)\geqslant\Delta$ is satisfied. In this region, $A(\omega,k,\Delta)\geqslant 0$
and, as a consequence, the quantities $\Pi_{00}$ and $\Pi$ have both the real and
imaginary parts. From the results of Ref.~\cite{46} generalized for the case $\mu\neq 0$,
after identical transformations, one obtains
\begin{linenomath}
\begin{eqnarray}
&&
\Pi_{00}\po=-\frac{2\alpha k^2}{cp^2(\omega,k)}\,\Phi_2(\omega,k,\Delta)+
\frac{4\alpha\hbar c^2\kp}{v_F^2}\left\{\int\limits_{\widetilde{D}}^{u_1}
du\,\Tw \right.
\nonumber\\
&&~~~\times
\left[1-\frac{1}{2c\kp}\sum_{\lambda=\pm 1} B_1(c\kp u+\lambda\omega)\right]
\label{eq19}\\
&&~~~~+\left.
\int\limits_{u_1}^{\infty}
du\,\Tw
\left[1-\frac{1}{2c\kp}\sum_{\lambda=\pm 1}\lambda B_1(c\kp u+\lambda\omega)\right]
\right\},
\nonumber
\end{eqnarray}
\end{linenomath}
where
\begin{linenomath}
\begin{equation}
\Phi_2(\omega,k,\Delta)=\Delta-\hbar c\kp\left[1+\frac{\Delta^2}{\hbar^2c^2p^2(\omega,k)}
\right]\,\left[{\rm arctanh}\frac{\Delta}{\hbar c\kp}+i\frac{\pi}{2}\right]
\label{eq20}
\end{equation}
\end{linenomath}
and the roots of the common denominator of the functions $B_1$ and $B_2$ are
\begin{linenomath}
\begin{equation}
u_{1,2}=\frac{1}{c\kp}\left[\omega\mp v_Fk\sqrt{A(\omega,k,\Delta)}\right].
\label{eq21}
\end{equation}
\end{linenomath}
These roots belonging to the integration domain $\widetilde{D}<u_1<u_2<\infty$
are obtained for $\lambda=-1$.

In a similar manner, $\Pi$ is given by
\begin{linenomath}
\begin{eqnarray}
&&
\Pi\po=\frac{2\alpha k^2}{c}\,\Phi_2(\omega,k,\Delta)+
\frac{4\alpha\hbar \omega^2\kp}{v_F^2}\left\{\int\limits_{\widetilde{D}}^{u_1}
du\,\Tw \right.
\nonumber\\
&&~~~\times
\left[1-\frac{c\kp}{2\omega^2}\sum_{\lambda=\pm 1} B_2(c\kp u+\lambda\omega)\right]
\label{eq22}\\
&&~~~~+\left.
\int\limits_{u_1}^{\infty}
du\,\Tw
\left[1-\frac{c\kp}{2\omega^2}\sum_{\lambda=\pm 1}\lambda B_2(c\kp u+\lambda\omega)\right]
\right\}.
\nonumber
\end{eqnarray}
\end{linenomath}
Note that the imaginary parts of $\Pi_{00}$ and $\Pi$ arise in the second integrals in
Eqs.~(\ref{eq19}) and (\ref{eq22}) when integrating with respect to $u$ from $u_1$ to $u_2$.

Let us continue and consider the region (\ref{eq14}). In this region, the analytic
continuation of the polarization tensor (\ref{eq4}) and (\ref{eq6}) to the real
frequency axis can be obtained by putting $i\xi_{E,l}=\omega$ with the appropriate
choice of the branch of the square root in denominator \cite{46}.
For this purpose, it is first convenient to
represent the real part in Eqs.~(\ref{eq4}) and (\ref{eq6}) as  half a sum of the
complex conjugated quantities. Then, the quantity $\Pi_{00}$ in the region (\ref{eq14})
of real frequency axis takes the form
\begin{linenomath}
\begin{eqnarray}
&&
\Pi_{00}\po=\frac{\alpha\hbar k^2}{\ktq}\Psi(D)+
\frac{4\alpha\hbar c^2\ktq}{v_F^2}\int\limits_{D}^{\infty}du\,\Ww
\nonumber\\
&&~~~\times\left\{1-\frac{1}{2}\sum_{\lambda=\pm 1}\lambda
\frac{1- u^2+2\lambda\gt u}{\left[1-u^2+2\lambda\gt u+
D^2+\widetilde{\gamma}^2(\omega,k)D^2\right]^{1/2}}\right\},
\label{eq23}
\end{eqnarray}
\end{linenomath}
where, in accordance with Eq.~(\ref{eq5}),

\begin{eqnarray}
&&
\ktq=\frac{1}{c}\sqrt{v_F^2k^2-\omega^2}, \qquad
\gt=\frac{\omega}{\sqrt{v_F^2k^2-\omega^2}},
\nonumber\\
&&
D\equiv D(\omega,k,\Delta)=\frac{\Delta}{\hbar c\ktq},\qquad
B(\omega,k,T_p)=\frac{\hbar c \ktq}{2k_BT_p}
\nonumber\\
&&
\Ww=\sum_{\kappa=\pm 1}\left[e^{B(\omega,k,T_p)u+\kappa\frac{\mu}{k_BT_p}}
+1\right]^{-1}\!\!\!.
\label{eq24}
\end{eqnarray}

In a similar way, the analytic continuation of the quantity $\Pi$  to
the region (\ref{eq14}) of real frequency axis is

\begin{eqnarray}
&&
\Pi\po={\alpha\hbar k^2}{\ktq}\Psi(D)+
\frac{4\alpha\hbar \ktq\omega^2}{v_F^2}\int\limits_{D}^{\infty}du\,\Ww
\nonumber\\
&&~~~\times\left\{1-\frac{1}{2}\sum_{\lambda=\pm 1}\lambda
\frac{[1-\lambda \widetilde{\gamma}^{-1}(\omega,k)u]^2-
[\widetilde{\gamma}^{-2}(\omega,k)+1]D^2}{\left[1-u^2+2\lambda\gt u+
D^2+\widetilde{\gamma}^2(\omega,k)D^2\right]^{1/2}}\right\}.
\label{eq25}
\end{eqnarray}

Let us discuss the obtained expressions for the polarization tensor of gapped and
doped graphene along the real frequency axis. First of all let us note that the
polarization tensor behaves differently depending on whether the inequality
$\Delta>2\mu$ or $\Delta\leqslant 2\mu$ is satisfied.

If $\Delta>2\mu$ the powers of exponents in the definitions of $\widetilde{w}$ and
$w$ in Eqs.~(\ref{eq16}) and (\ref{eq24}) are positive even at the lower integration
limits $u=\widetilde{D}$ and $u=D$ in Eqs.~(\ref{eq15}) and (\ref{eq17}),
(\ref{eq19}) and (\ref{eq22}), (\ref{eq23}) and (\ref{eq25}).  This is true because
from Eqs.~(\ref{eq16}) and (\ref{eq24}) one obtains

\begin{equation}
\widetilde{B}\widetilde{D}=BD=\frac{\Delta}{2k_BT_p}>\frac{\mu}{k_BT_p}.
\label{eq26}
\end{equation}

In this case, Eqs.(\ref{eq16}) and (\ref{eq24}) lead to
\begin{linenomath}
\begin{equation}
\lim_{T_p\to 0}\Ww=\lim_{T_p\to 0}\Tw=0.
\label{eq27}
\end{equation}
\end{linenomath}
As a result, the polarization tensor at zero temperature defined as
\begin{linenomath}
\begin{eqnarray}
&&
\Pi_{00}\zo=\lim_{T_p\to 0}\Pi_{00}\po,
\nonumber \\
&&
\Pi\zo=\lim_{T_p\to 0}\Pi\po
\label{eq28}
\end{eqnarray}
\end{linenomath}
takes a much simplified form.  It is equal to the first terms in Eqs.~(\ref{eq15})
and (\ref{eq17}),
\begin{linenomath}
\begin{eqnarray}
&&
\Pi_{00}\zo=-\frac{2\alpha k^2}{cp^2(\omega,k)}\Phi_1(\omega,k,\Delta),
\nonumber \\
&&
\Pi\zo=\frac{2\alpha k^2}{c}\Phi_1(\omega,k,\Delta),
\label{eq29}
\end{eqnarray}
\end{linenomath}
in the plasmonic region (\ref{eq13}) under the condition $\hbar cp<\Delta$,
to the first terms in Eqs.~(\ref{eq19})
and (\ref{eq22}),
\begin{linenomath}
\begin{eqnarray}
&&
\Pi_{00}\zo=-\frac{2\alpha k^2}{cp^2(\omega,k)}\Phi_2(\omega,k,\Delta),
\nonumber \\
&&
\Pi\zo=\frac{2\alpha k^2}{c}\Phi_2(\omega,k,\Delta),
\label{eq30}
\end{eqnarray}
\end{linenomath}
in the plasmonic region (\ref{eq13}) under the condition $\hbar cp\geqslant\Delta$,
and to the first terms in Eqs.~(\ref{eq23})
and (\ref{eq25}),
\begin{linenomath}
\begin{eqnarray}
&&
\Pi_{00}\zo=\frac{\alpha\hbar k^2}{\ktq}\Psi(D),
\nonumber \\
&&
\Pi\zo={\alpha\hbar k^2}{\ktq}\Psi(D),
\label{eq31}
\end{eqnarray}
\end{linenomath}
in the region (\ref{eq14}).

Thus, if the inequality $\Delta>2\mu$ is satisfied, the polarization tensor at zero
temperature (\ref{eq28}) does not depend on the chemical potential $\mu$.

It is easily seen that under the opposite inequality $\Delta<2\mu$ there are the
intervals in the integration domains where the powers of exponents in the
definitions of $\widetilde{w}$ and $w$ in Eqs.~(\ref{eq16}) and (\ref{eq24}) with
$\kappa=-1$ are negative. Thus, in the plasmonic region (\ref{eq13}) the power
of exponent in  $\widetilde{w}$ with $\kappa=-1$ is negative within the interval

\begin{equation}
\widetilde{D}<u<u_0=\frac{2\mu}{\hbar c\kp}.
\label{eq32}
\end{equation}

In a similar way, in the region (\ref{eq14}) the power
of exponent in  ${w}$ with $\kappa=-1$ is negative in the interval

\begin{equation}
{D}<u<\tilde{u}_0=\frac{2\mu}{\hbar c\ktq}.
\label{eq33}
\end{equation}

In both cases it holds
\begin{linenomath}
\begin{equation}
\lim_{T_p\to 0}\Ww=\lim_{T_p\to 0}\Tw=1.
\label{eq34}
\end{equation}
\end{linenomath}
As a result, not only the first but all terms in Eqs.~(\ref{eq15}) and (\ref{eq17}),
(\ref{eq19}) and (\ref{eq22}), (\ref{eq23}) and (\ref{eq25}) contribute to the
polarization tensor at zero temperature defined in Eq.~(\ref{eq28}).

By way of example, in the region (\ref{eq14}) the polarization tensor of graphene with
the mass gap and chemical potential satisfying the condition $\Delta<2\mu$ takes the
following form at $T=0$:
\begin{linenomath}
\begin{eqnarray}
&&
\Pi_{00}\zo=\frac{\alpha\hbar k^2}{\ktq}\Psi(D)+\frac{4\alpha\hbar c^2\ktq}{v_F^2}
\nonumber \\
&&~~
\times\left\{\frac{2\mu-\Delta}{\hbar c\ktq}-\frac{1}{2}
\int\limits_{D}^{\tilde{u}_0}du\sum_{\lambda=\pm 1}\lambda
\frac{1- u^2+2\lambda\gt u}{\left[1-u^2+2\lambda\gt u+
D^2+\widetilde{\gamma}^2(\omega,k)D^2\right]^{1/2}}\right\},
\nonumber\\
&&
\Pi\zo={\alpha\hbar k^2}{\ktq}\Psi(D)+\frac{4\alpha\hbar \ktq\omega^2}{v_F^2}
\label{eq35}\\
&&~~
\times\left\{\frac{2\mu-\Delta}{\hbar c\ktq}-\frac{1}{2}
\int\limits_{D}^{\tilde{u}_0}du\sum_{\lambda=\pm 1}\lambda
\frac{[1-\lambda \widetilde{\gamma}^{-1}(\omega,k)u]^2-
[\widetilde{\gamma}^{-2}(\omega,k)+1]D^2}{\left[1-u^2+2\lambda\gt u+
D^2+\widetilde{\gamma}^2(\omega,k)D^2\right]^{1/2}}\right\},
\nonumber
\end{eqnarray}
\end{linenomath}
where the upper integration limit $\tilde{u}_0$ is defined in Eq.~(\ref{eq33}).

Thus, under the
condition $\Delta<2\mu$ the polarization tensor at zero temperature depends on the
chemical potential $\mu$ through the terms in the figure brackets in Eq.~(\ref{eq35})
and similar terms surviving at $T=0$ in Eqs.~(\ref{eq15}), (\ref{eq17}) and
(\ref{eq19}), (\ref{eq22}) valid in the plasmonic region.

Note that in the plasmonic region (\ref{eq13}) the expression for the polarization
tensor at zero temperature (\ref{eq30}), found under the opposite condition
$2\mu<\Delta\leqslant\hbar cp$ contains the pure imaginary part
for some values of $\omega$ and $k$ satisfying the inequality $\hbar cp\geqslant\Delta$.
This is in line with
the fact that additional contributions to the zero-temperature
polarization tensor arising under the condition $\Delta<2\mu$
 also contain the imaginary parts. It is only a quantitative difference.

Another situation arises in the region (\ref{eq14}) where $k>\omega/v_F$.
Here, under the condition $\Delta>2\mu$, the polarization tensor at zero temperature
(\ref{eq31}) is real, but under the condition $\Delta<2\mu$ it gains some imaginary
part depending on $\mu$
for some values of $\omega$ and $k$. This leads to important qualitative
differences between the cases $\Delta>2\mu$ and $\Delta<2\mu$, which are demonstrated
in Section~4 by the results of numerical computations.

\section{Computational Results for the Nonequilibrium Casimir-Polder Force}

In this section, the nonequilibrium Casimir-Polder force $\nfeq$ from Eq.~(\ref{eq8})
between nanoparticles and a silica plate coated with a gapped and doped graphene sheet
is computed using the formalism presented in Sections 2 and 3. As in Section 2,
the temperatures of nanoparticles and of the environment are assumed to be equal
to $T_E=300~$K, whereas the temperature of the graphene-coated plate is either lower,
$T_p=77~$K, or higher, $T_p=500~$K, than that of the environment.
All computations were performed for typical values of the mass-gap parameter
$\Delta=0.1$ and 0.2~eV   and the chemical potential $\mu=0$, 0.075 and 0.15~eV
of the graphene coating within the separation region from 200~nm to $2~\upmu$m.

Numerical computations of the first contribution $\feq$ to $\nfeq$ given by
 Eq.~(\ref{eq1}) were performed along the imaginary frequency axis as discussed in
 Section~2 but with $T_p\neq T_E$. The computational results for the second
 contribution $F_r^{\rm SiO_2}$ to $\nfeq$ were obtained by
 Eq.~(\ref{eq9}) using the polarization tensor of graphene along the real
 frequency axis defined for different values of $\omega$ and $k$
in Eqs.~(\ref{eq15}) and (\ref{eq17}),
(\ref{eq19}) and (\ref{eq22}), (\ref{eq23}) and (\ref{eq25}).
Numerical data for the real and imaginary parts of the dielectric permittivity
of SiO${}_2$ are taken from Ref.~\cite{81} (see also Ref.~\cite{69} for the
graphical representation of these data).
Note that numerical computations of $\nfeq$ are much more complicated than those for
$\feq$ made along the imaginary frequency axis owing to the quick variation of the
integrands. They were performed using the program utilizing the Gauss-Kronrod and
double-exponential quadrature methods for numerical integration from the GNU
scientific Library \cite{85} and Boost ${\rm C}^{++}$ Libraries \cite{86}.
The program was written in the ${\rm C}^{++}$  programming language. It utilizes
the OpenMP Library \cite{87} for parallelism. The numerical computations were
performed on a supercomputer of the Peter the Great Saint Petersburg Polytechnic
University. Sufficiently high precision of the computational results  was reached
using the Boost Multiprecision Library \cite{88}.

\begin{figure}[H]
\centerline{\hspace*{-2.7cm}
\includegraphics[width=5in]{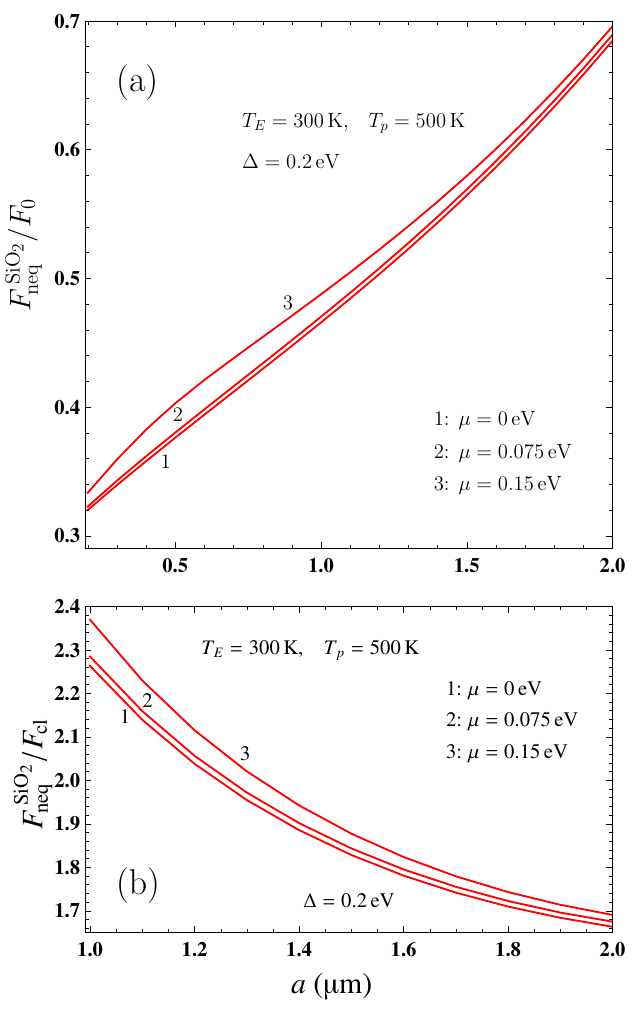}}
\caption{\label{fg3}
The nonequilibrium Casimir-Polder force between a nanoparticle and a graphene-coated
SiO${}_2$ plate kept at $T_p=500~$K normalized to (a) the zero-temperature Casimir-Polder
force from an ideal metal plane and (b) classical limit of the same force at
$T_E=300~$K is shown as the function of separation
for a graphene coating with the mass gap $\Delta=0.2~$eV and chemical potential
$\mu=0$, 0.075, and 0.15~eV  by the lines labeled 1, 2, and 3, respectively.
}
\end{figure}

We begin with the case when the SiO${}_2$ plate coated with gapped and doped graphene
is heated up to 500~K  whereas the temperature of nanoparticles and
of the environment here and below is always $T_E=300~$K. The mass-gap parameter of
graphene coating is $\Delta=0.2~$eV, whereas the chemical potential can be equal to
zero or take the values $\mu=0.075~$eV and 0.15~eV. In Figure~\ref{fg3} the
computational results  for the nonequilibrium Casimir-Polder force normalized to
(a) the zero-temperature Casimir-Polder force $F_0$ acting on a nanoparticle from
an ideal metal plane and (b) the classical limit $F_{\rm cl}$ of the same force
at $T_p=T_E=300~$K [see Eq.~(\ref{eq7})]  are shown as the function of separation
by the lines labeled 1, 2, and 3 for graphene coating with $\mu=0$, 0.075, and
0.15~eV, respectively.

{}From Figure~\ref{fg3} it is seen that with increasing chemical potential the
magnitude of the nonequilibrium Casimir-Polder force increases. Thus, in this
respect the nonequilibrium force behaves in the same way as the equilibrium one.
Note that the lines 1 and 2, for which $2\mu<\Delta=0.2~$eV, are rather close to each
other at all separations. The line 3, for which $2\mu>\Delta$, is further apart
from the lines 1 and 2 at short separations but approach them at larger separations.
Thus, for a heated graphene-coated plate there is no qualitative difference between
the cases $2\mu<\Delta$ and $2\mu>\Delta$.

Now we consider the situation when the graphene-coated plate is cooled to $T_p=77~$K
preserving unchanged all the other parameters of graphene coating listed in
Figure~\ref{fg3}. The computational results for the normalized nonequilibrium
Casimir-Polder force are presented as a function of separation in Figure~\ref{fg4}
using the same notations  for all lines as in Figure~\ref{fg3}.

As is seen in Figure~\ref{fg4}, the case of a cooled graphene-coated plate differ
greatly from the case of a heated one. In the case of graphene-coated plate kept
at $T_p=77~$K, an impact of the chemical potential on the nonequilibrium
Casimir-Polder force  essentially depends on the relationship between $2\mu$
and $\Delta$.

\begin{adjustwidth}{-\extralength}{0cm}
\begin{figure}[H]
\centerline{\hspace*{-2.7cm}
\includegraphics[width=7in]{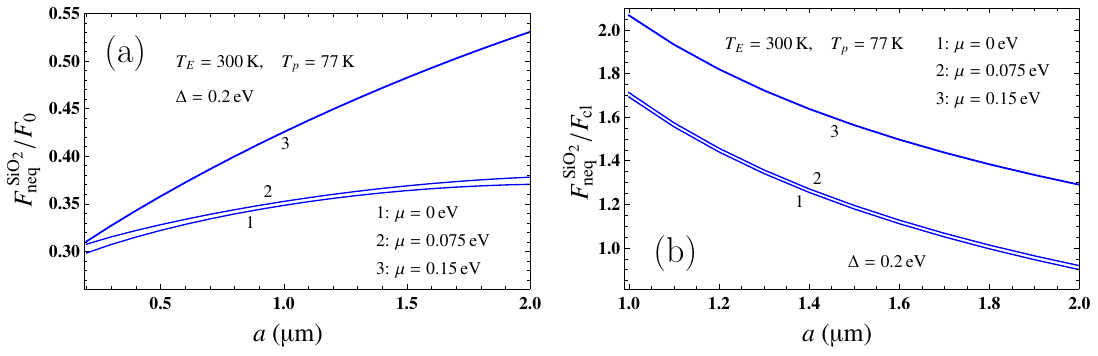}}
\caption{\label{fg4}
The nonequilibrium Casimir-Polder force between a nanoparticle and a graphene-coated
SiO${}_2$ plate at $T_p=77~$K normalized to (a) the zero-temperature Casimir-Polder
force from an ideal metal plane and (b) classical limit of the same force at
$T_E=300~$K is shown as the function of separation
for a graphene coating with the mass gap $\Delta=0.2~$eV and chemical potential
$\mu=0$, 0.075, and 0.15~eV  by the lines labeled 1, 2, and 3, respectively.
}
\end{figure}
\end{adjustwidth}

If $2\mu<\Delta$ (this holds for the line labeled 2) the impact of nonzero $\mu$
on the force value  is rather moderate. If, however, $2\mu>\Delta$ (line 3), this
leads to significant deviation of the force value from that computed at $\mu=0$
(line 1) which increases with increasing separation between nanoparticles and a
graphene-coated plate. The enhancement of nonequilibrium force for the graphene
coating with $2\mu>\Delta$ is explained by the increasing role of the polarization
tensor at $T=0$, which has a nonzero imaginary part in this case, and decrease of
the thermal correction to it (see the discussion in the end of Section 3).
By and large, although the nonequilibrium Casimir-Polder force increases with
increasing $\mu$ for both the heated and cooled graphene-coated plate, in the
latter case the impact  of $\mu$ on the force value is much stronger.

We consider now the SiO${}_2$ plate coated with the graphene sheet possessing
the smaller mass-gap parameter $\Delta=0.1~$eV and the moderate chemical potential
$\mu=0.075~$eV. In this case, the condition $2\mu>\Delta$ is satisfied.
The nonequilibrium Casimir-Polder force acting on nanoparticles in such a configuration
was computed at three different temperatures of the graphene-coated plate:
$T_p=77~$K (the cooled plate), $T_p=300~$K (the situation of the thermal equilibrium),
and $T_p=500~$K (the heated plate).

The computational results for the Casimir-Polder force normalized to (a) the force
$F_0$ from an ideal metal plane at zero temperature and (b) the classical force
$F_{\rm cl}$ from an ideal metal plane at $T_p=T_E=300~$K are shown in Figure~\ref{fg5}
as the function of separation by the lines labeled 1, 2, and 3 for temperatures of the
graphene-coated plate $T_p=77~$K, 300~K, and 500~K, respectively.
As is seen in Figure~\ref{fg5}, the Casimir-Polder force at all separations increases
with increasing temperature. This increase is smaller when the temperature of the
graphene-coated plate $T_p$ is below the temperature of the environment $T_E$ and larger
when $T_P>T_E$. In doing so, the impact of temperature on the force value increases
with increasing separation.

It is interesting to consider the role of the mass-gap parameter. For this purpose,
the same configuration at the same three temperatures of the graphene-coated plate
was considered with the larger mass-gap parameter of the graphene sheet
$\Delta=0.2~$eV. In this case the opposite inequality $2\mu=0.15~\mbox{eV}<\Delta$
is satisfied.

\begin{adjustwidth}{-\extralength}{0cm}
\begin{figure}[H]
\centerline{\hspace*{-2.7cm}
\includegraphics[width=7in]{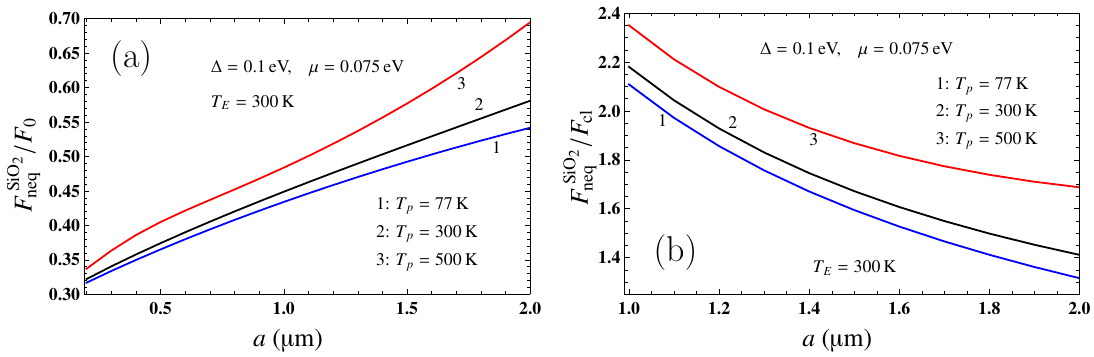}}
\caption{\label{fg5}
The Casimir-Polder force between a nanoparticle and a  SiO${}_2$ plate coated
by a graphene sheet with the mass-gap parameter $\Delta=0.1~$eV and chemical
potential $\mu=0.075~$eV
normalized to (a) the  Casimir-Polder force from an ideal metal plane at
zero temperature and (b) classical limit of the same force at
$T_E=300~$K is shown as the function of separation
for a graphene-coated plate temperatures $T_p=77~$K (thermal nonequilibrium),
$T_p=300~$K (thermal equilibrium), and $T_p=500~$K (thermal nonequilibrium)
by the lines 1, 2, and 3, respectively.
}
\end{figure}
\end{adjustwidth}

The computational results for the nonequilibrium (lines 1 and 3) and equilibrium
(line 2) Casimir-Polder force normalized to (a) $F_0$ and (b) $F_{\rm cl}$ are
shown on Figure~\ref{fg6} by the lines labeled 1, 2, and 3 for temperatures of the
graphene-coated plate $T_p=77~$K, 300~K, and 500~K, respectively.
Similar to Figure~\ref{fg5}, the Casimir-Polder force increases with increasing
temperature and this increase becomes larger at larger separations.

\begin{adjustwidth}{-\extralength}{0cm}
\begin{figure}[H]
\centerline{\hspace*{-2.7cm}
\includegraphics[width=7in]{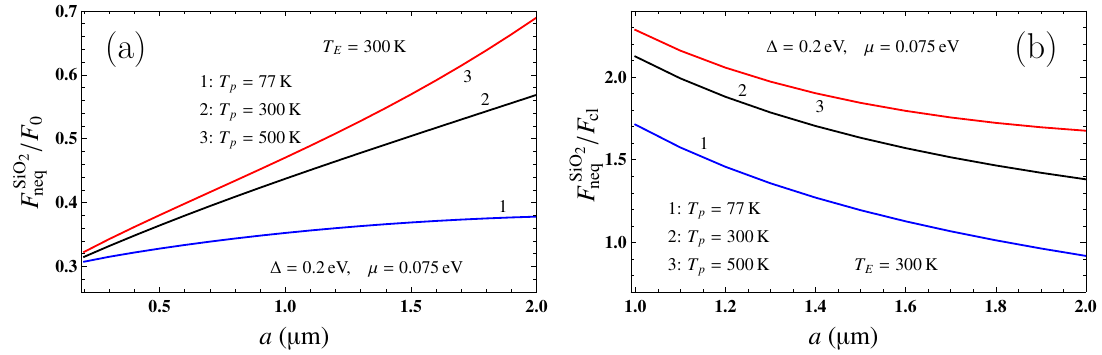}}
\caption{\label{fg6}
The Casimir-Polder force between a nanoparticle and a  SiO${}_2$ plate coated
by a graphene sheet with the mass-gap parameter $\Delta=0.2~$eV and chemical
potential $\mu=0.075~$eV
normalized to (a) the  Casimir-Polder force from an ideal metal plane at
zero temperature and (b) classical limit of the same force at
$T_E=300~$K is shown as the function of separation
for a graphene-coated plate temperatures $T_p=77~$K (thermal nonequilibrium),
$T_p=300~$K (thermal equilibrium), and $T_p=500~$K (thermal nonequilibrium)
by the lines 1, 2, and 3, respectively.
}
\end{figure}
\end{adjustwidth}

By comparing the force values in  Figure~\ref{fg5} and  Figure~\ref{fg6}, it is
seen that at 77~K (the lines labeled 1 in both figures) the value of the
nonequilibrium Casimir-Polder force for the graphene sheet with $\Delta=0.1~$eV
(Figure~\ref{fg5}) is much larger than that for the graphene sheet with $\Delta=0.2~$eV
(Figure~\ref{fg6}). This is because under the condition $2\mu>\Delta$ satisfied for
the configuration considered in Figure~\ref{fg5} the impact of the polarization
tensor at $T=0$ is significantly increased as discussed in the end of Section~3.
As to the force values shown by the lines 2 and 3, in Figure~\ref{fg6}, where
$2\mu<\Delta$, they are only slightly smaller than those in Figure~\ref{fg5}.

Finally, we consider the same configuration, as in Figure~\ref{fg6}, i.e., the
SiO${}_2$ plate coated by a graphene sheet with the mass-gap parameter
$\Delta=0.2~$eV, but with the twice as large chemical potential $\mu=0.15~$eV.
In this case, the condition $2\mu>\Delta$ is satisfied at the cost of sufficiently
large $\mu$, but not at the cost of relatively small $\Delta$ as in Figure~\ref{fg5}.
The computational results normalized to (a) $F_0$ and (b) $F_{\rm cl}$ are
shown on Figure~\ref{fg7} as the function of separation by the lines 1 and 3 for
the nonequilibrium Casimir-Polder force computed at  $T_p=77~$K and 500~K,
respectively, and by the line 2 for the equilibrium force computed at  $T_p=300~$K.

\begin{adjustwidth}{-\extralength}{0cm}
\begin{figure}[H]
\centerline{\hspace*{-2.7cm}
\includegraphics[width=7in]{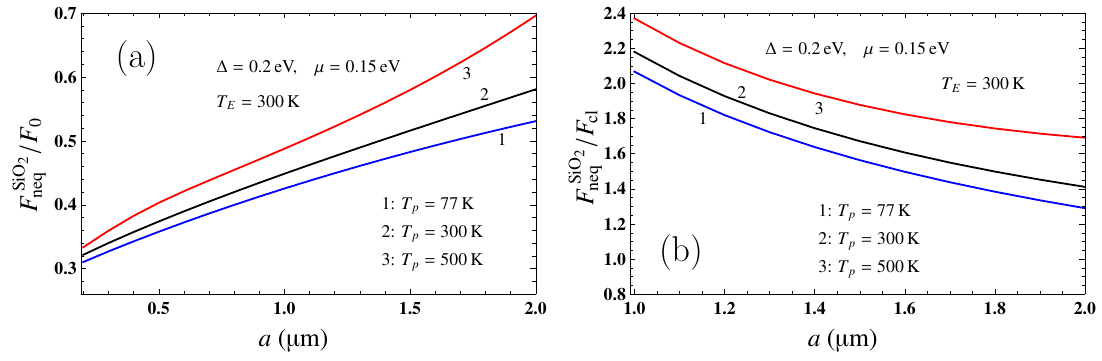}}
\caption{\label{fg7}
The Casimir-Polder force between a nanoparticle and a  SiO${}_2$ plate coated
by a graphene sheet with the mass-gap parameter $\Delta=0.2~$eV and chemical
potential $\mu=0.15~$eV
normalized to (a) the  Casimir-Polder force from an ideal metal plane at
zero temperature and (b) classical limit of the same force at
$T_E=300~$K is shown as the function of separation
for a graphene-coated plate temperatures $T_p=77~$K (thermal nonequilibrium),
$T_p=300~$K (thermal equilibrium), and $T_p=500~$K (thermal nonequilibrium)
by the lines 1, 2, and 3, respectively.
}
\end{figure}
\end{adjustwidth}

Similar to the previous figures, the force values in Figure~\ref{fg7} increase with
increasing temperature, and this increase is more pronounced at larger separations
between nanoparticles and a graphene-coated plate. There is only a minor impact of
the increased chemical potential on the lines 2 and 3 relevant to the equilibrium
situation and to the heated graphene-coated plate, respectively. These lines
demonstrate only slightly larger force values in comparison with those in Figure~\ref{fg6}.
One can conclude that at sufficiently high temperature the polarization tensor is
almost independent on whether $2\mu>\Delta$ or $2\mu<\Delta$ although the
zero-temperature contribution and the thermal correction to it are different under these
conditions. As to the line 1 in Figure~\ref{fg7}, computed at $T_p=77~$K, it
demonstrates much larger values of the nonequilibrium Casimir-Polder force than the
line 1 in Figure~\ref{fg6}. This effect is explained by the fact that at $2\mu>\Delta$
the polarization tensor at zero temperature takes the relatively large value.
Specifically, it has a nonzero imaginary part in the region (\ref{eq14}) (see a discussion
in the end of Section~3). For a cooled plate just this region contributes to the increase
of the force value.

\section{Discussion}

As was shown in Section~1, the Casimir-Polder interaction of nanoparticles with
material substrates and, especially, with graphene-coated plates attracts much
attention in fundamental physics and finds  applications in nanotechnology.
Because of this, it is important to reliably predict the strength of this
interaction and its dependence on all relevant parameters. Graphene is a novel
material and, to a first approximation, its response to the electromagnetic field
was described by using different phenomenological approaches, such as Kubo
theory, 2D Drude model etc. (see Refs.~\cite{2,3,4}) for a review).

Development of the formalism of the polarization tensor \cite{44,45,46,47} has
made it possible to calculate the equilibrium Casimir and Casimir-Polder forces
in graphene systems in the framework of the Dirac model on the basis of first
principles of thermal quantum field theory at nonzero temperature.
However, application of the same formalism to the nonequilibrium forces, although
rather straightforward theoretically, presents serious computational difficulties.
The reason is that the equilibrium Casimir and Casimir-Polder forces are
calculated as the sum over the discrete pure imaginary Matsubara frequencies,
whereas computation of the nonequilibrium forces is unavoidably involved with
the integration of rapidly varying functions along the real frequency axis
\cite{50,51,52,53,54,55,56,57,58,59,60,61}, which was always challenging in the
Casimir physics.

This difficulty was overcome step by step. At first, the nonequilibrium
Casimir-Polder force between nanoparticles and a freestanding sheet of a
pristine graphene was considered \cite{65}. This case is interesting from the
theoretical point of view, but does not reflect realistic experimental
situations. Next, the impact of the mass gap, which exists in real graphene
sheets, on the nonequilibrium force was investigated \cite{68}.
 Among other things, the mass gap necessary arises in graphene sheets deposited
 on some substrate. Because of this, the obtained results were generalized for
 the case of a graphene-coated dielectric plate \cite{69} as usually takes place
 in experimental situations. All these computations demanded the use of
 a supercomputer.

 One more important parameter, which was as yet disregarded, is the chemical
 potential reflecting the concentration of foreign atoms in a graphene sheet.
 As was shown previously, just the relationship between the mass gap and chemical
 potential determines the behavior of the Casimir and Casimir-Polder forces in
 graphene systems \cite{38,40,42}. In Ref.~\cite{73a} the impact of chemical
 potential on the nonequilibrium Casimir force between two parallel
 graphene-coated plates was investigated, but with disregarded mass gap of
 graphene and using the spatially local model of its electromagnetic response.

 The formalism and computational results presented in this paper take into account
 all the parameters important for an adequate description of the nonequilibrium
 Casimir-Polder interaction between nanoparticles and a graphene-coated plate in
 the experimental situation, i.e., the mass gap and chemical potential of a
 graphene sheet, the dielectric permittivity of the plate material and their
 temperature. As expected from the experience obtained  when studying the
graphene systems in thermal equilibrium, the nonequilibrium Casimir-Polder force
between nanoparticles and a graphene-coated substrate essentially  depends on the
relationship between the energy gap and twice the chemical
 potential of the graphene coating. This opens opportunities for high-reliability
 predictions of the Casimir-Polder forces in graphene systems out of thermal
 equilibrium for a subsequent comparison with the experimental data.

\section{Conclusions}

To conclude, in the foregoing we have investigated the Casimir-Polder
force between spherical nanoparticles and a graphene-coated SiO$_2$
plate in out-of-thermal-equilibrium conditions. The response of graphene
to the electromagnetic field was described by the spatially nonlocal
polarization tensor taking into account both the mass gap and chemical
potential of the graphene coating, which was found in the framework of
the Dirac model. The plate material was characterized by the
frequency-dependent dielectric permittivity obtained from the tabulated
optical data for the complex index of refraction of silica glass.

By computing the nonequilibrium Casimir-Polder force on a supercomputer,
the following results were obtained. With increasing chemical potential
$\mu$ of the graphene coating, the nonequilibrium Casimir-Polder force
increases for both heated and cooled graphene-coated plate. An impact
of the chemical potential on the force value is stronger when the
graphene-coated plate is cooled, as compared to the environment temperature,
and weaker when it is heated. An impact of the mass gap $\Delta$ of the
graphene coating is also more pronounced for a cooled graphene-coated plate.
In this case, the character of the nonequilibrium Casimir-Polder force
essentially depends on the relationship between the values of $\Delta$ and
$2\mu$. An impact of the chemical potential on the force value is rather
moderate if $2\mu<\Delta$, but if $2\mu>\Delta$ the magnitude of the
nonequilibrium force becomes much larger than that computed for a graphene
coating with $\mu=0$. This effect is explained by the increasing role
of the polarization tensor of graphene defined at zero temperature.

The magnitude of the nonequilibrium Casimir-Polder force increases with
increasing temperature of the graphene-coated plate, all other conditions
being equal. This increase is weaker and stronger if the temperature of
the graphene-coated plate is lower and higher than that of the environment,
respectively. In all cases the impact of temperature on the force value
becomes stronger with increasing separation between nanoparticles and a
graphene-coated plate.

To conclude, the revealed dependencies of the nonequilibrium Casimir-Polder
force between nanoparticles and a graphene-coated plate from the mass-gap
parameter, chemical potential, and temperature provide a way to control
the force value in the graphene-based nanotechnological devices of next
generations.

\vspace{6pt}

\funding{G.L.K.and V.M.M. were partially funded by the
Ministry of Science and Higher Education of Russian Federation
("The World-Class Research Center: Advanced Digital Technologies,"
contract No. 075-15-2022-311 dated April 20, 2022). The research
of V.M.M. was partially carried out in accordance with the Strategic
Academic Leadership Program "Priority 2030" of the Kazan Federal
University. }

\begin{adjustwidth}{-\extralength}{0cm}

\reftitle{References}

\end{adjustwidth}
\end{document}